\documentclass[12pt]{article}
\usepackage{graphicx}
\usepackage{lineno}
\usepackage{url}
\usepackage{amsmath}
\usepackage{subcaption}
\usepackage{subfig}

\newcommand\pubnumber{ATL-PHYS-PROC-2014-083}
\newcommand\pubdate{25 July 2014}

\def\Title#1{\begin{center} {\Large #1 } \end{center}}
\def\Author#1{\begin{center}{ \sc #1} \end{center}}
\def\Address#1{\begin{center}{ \it #1} \end{center}}

\newcommand\pubblock{\rightline{\begin{tabular}{l} \pubnumber\\
         \pubdate  \end{tabular}}}
\newenvironment{Abstract}{\begin{quotation}  }{\end{quotation}}
\newenvironment{Presented}{\begin{quotation} \begin{center} 
             PRESENTED AT\end{center}\bigskip 
      \begin{center}\begin{large}}{\end{large}\end{center} \end{quotation}}

\begin{document}

\begin{titlepage}
\pubblock

\vfill
\Title{Higgs and new physics at high energy}
\vfill
\Author{C.A. Solans\footnote{On behalf of the ATLAS and CMS Collaborations.}}
\Address{CERN, Geneve 23, CH-1211, Switzerland}
\vfill
\begin{Abstract}
The observation of a new particle in the search for the Standard Model (SM) Higgs boson at the LHC, reported by the ATLAS and CMS collaborations, is a milestone in the quest to understand electroweak symmetry breaking.
The evidence at the level of 5 $\sigma$ for a Higgs boson-like particle has been published by both experiments after a preliminary analysis of the data from the LHC Run-I. 
Precision measurements of the new particle are of critical importance. This document reviews the mass and spin measurement, the couplings scale factor measurements and the limits on new physics derived from these results.
\end{Abstract}
\vfill
\begin{Presented}
2014 Flavor Physics and CP Violation (FPCP-2014), Marseille, France, May 26-30 2014
\end{Presented}
\vfill
\end{titlepage}
\def\thefootnote{\fnsymbol{footnote}}
\setcounter{footnote}{0}

\section{Introduction}
The ATLAS~\cite{AtlasDetector} and CMS~\cite{CmsDetector} experiments reported the observation of a new particle in the search for the Standard Model (SM) Higgs boson at the LHC at the level of 5~$\sigma$ after a preliminary analysis of the data from proton-proton collisions at $\sqrt{s}$ = 7 and 8~TeV recorded in the years 2011 and 2012~\cite{HiggsDiscoveryAtlas,HiggsDiscoveryCms}.
The analysis is based on an integrated luminosity of 25~fb$^{-1}$, which represents 90\% of that delivered by the LHC.
In the following the mass and spin measurement is reviewed, as well as the couplings scale factor measurements 
and the limits on new physics derived from these results~\cite{ATL-CONF-2013-014,ATLAS-CONF-2014-009,CMS-PAS-HIG-13-002,CMS-PAS-HIG-13-005,CMS-HIG-13-033}.

\section{Higgs mass and spin measurement}

\begin{figure}[ht!]
\centering
\includegraphics[width=7cm]{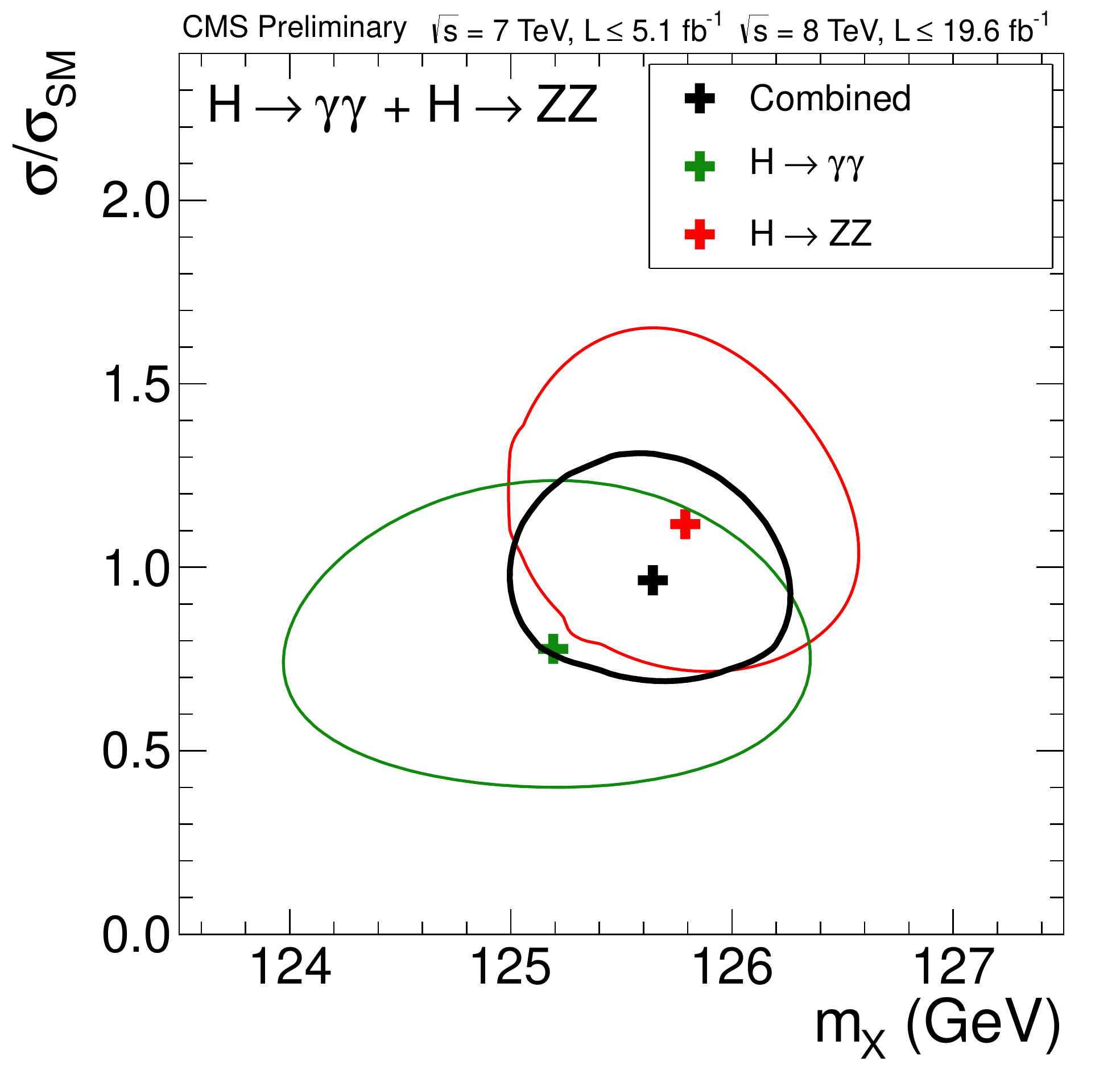}
\includegraphics[width=6.5cm, trim=0 -30 0 0, clip]{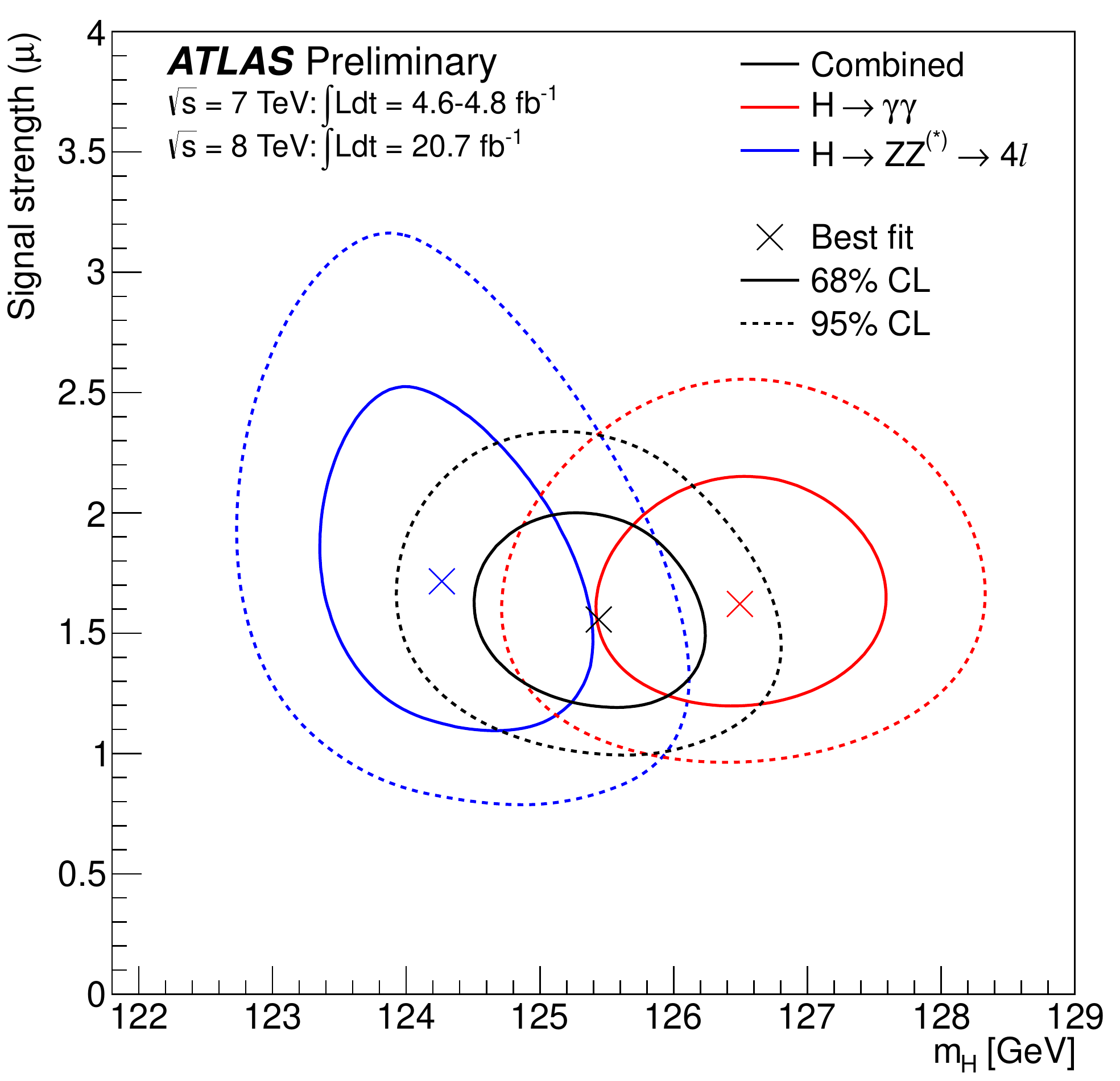}
\caption{\label{fig:MassMeasurement}
Confidence level intervals in the ($\mu$,$m_H$) plane for the $H\!\rightarrow\!\gamma\gamma$ and $H\!\rightarrow\!ZZ^*\!\rightarrow\!4\ell$ channels and their combination, including all systematic uncertainties for the ATLAS~\cite{ATLAS-CONF-2014-009} and CMS~\cite{CMS-PAS-HIG-13-005} experiments.}
\end{figure}

The mass is measured independently by both experiments in the $H\!\rightarrow\!\gamma\gamma$ and $H\!\rightarrow\!ZZ^*\!\rightarrow\!4\ell$ channels. Both experiments combine the data from both channels to measure the mass of the Higgs, using the invariant mass of the two photons and the four leptons respectively, based on a profile likelihood ratio~\cite{ProfileLikelihood} test statistic where the Higgs boson mass $m_{H}$ is the parameter of interest and all other uncertainties are nuisance parameters.
The Higgs mass is $m_H=125.5\pm{0.2}(\text{stat})^{+0.5}_{-0.4}(\text{syst})$ GeV for ATLAS and $m_H=125.7\pm{0.3}(\text{stat})\pm{0.3}(\text{syst})$ GeV for CMS. 
The measured mass value is close to 125.5~GeV as can be seen in Figure~\ref{fig:MassMeasurement}, that shows the contour plots for the signal strength as a function of the mass hypothesis.
The combined measurement is in the 68\% confidence interval for both channels in CMS and in the 95\% confidence interval for ATLAS. 
Note there is a difference between the individual Higgs mass measurements in ATLAS, which results from the uncertainty in the measurement of electrons and photons and is correlated between the two channels.
However this mass difference ($\Delta m_H = m^{\gamma\gamma}_{H} - m^{4\ell}_{H}$) is compatible with zero at the level of 2.4~$\sigma$. 

\begin{figure}[ht!]
\centering
\includegraphics[height=7.5cm, trim=15 0 15 0, clip]{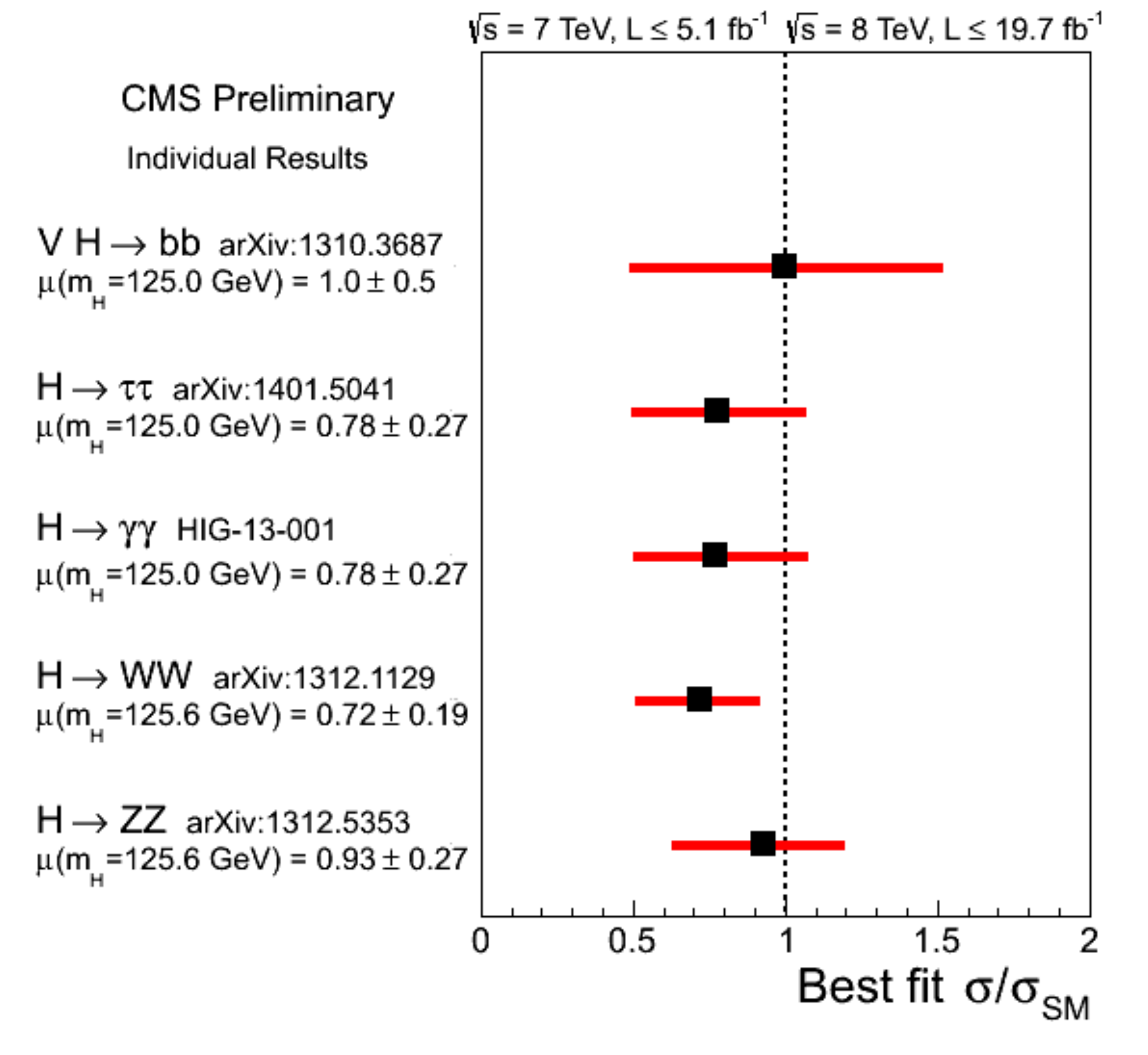}
\includegraphics[height=8.1cm, trim=10 0 5 0, clip]{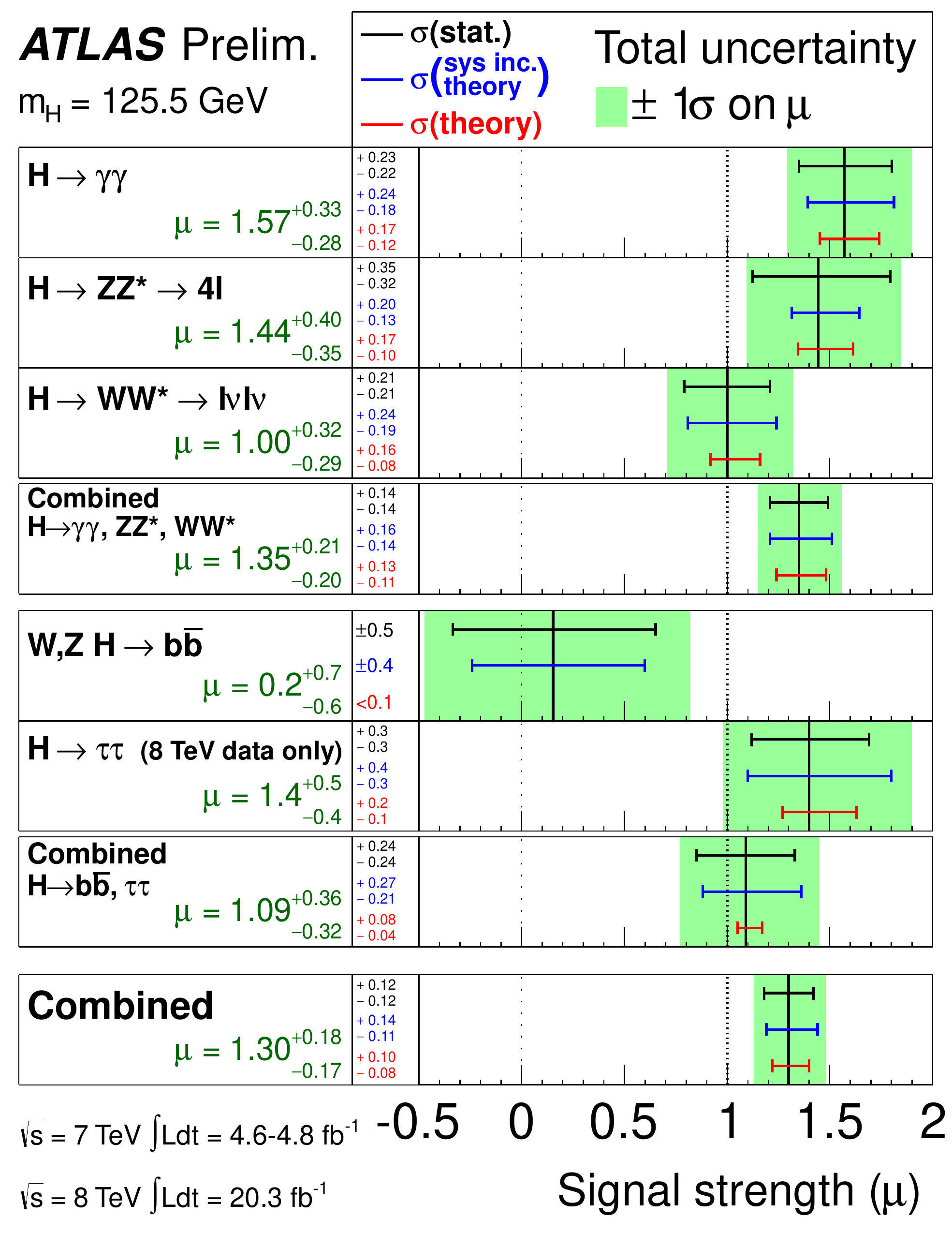}
\caption{Higgs signal strength for different decay channels measured by the ATLAS~\cite{ATLAS-CONF-2014-009} and CMS~\cite{CMS-Higgs-Physics-Results} experiments.\label{fig:SignalStrength}}
\end{figure}

The signal strength being the ratio of the measured cross section to the one predicted by the SM ($\mu=\sigma/\sigma_{SM}$) \cite{YellowReport2013} is slightly higher in ATLAS due to an excess of events observed in both channels, but compatible with the one measured by CMS.
Likewise, the signal strength observed in the different single channel searches ($H\!\rightarrow\!\gamma\gamma$, $H\!\rightarrow\!\tau\tau$, $H\!\rightarrow\!WW$, $H\!\rightarrow\!ZZ$, $VH\!\rightarrow\!b\bar{b}$) are compatible across experiments, as shown in Figure~\ref{fig:SignalStrength}, even for the $VH\rightarrow b\bar{b}$ search where the central values are quite different, but the contours overlap.
The signal strength is an important parameter, large deviations of this quantity from unity could indicate evidence for new physics.

\begin{figure}[ht!]
\centering
\includegraphics[height=7cm]{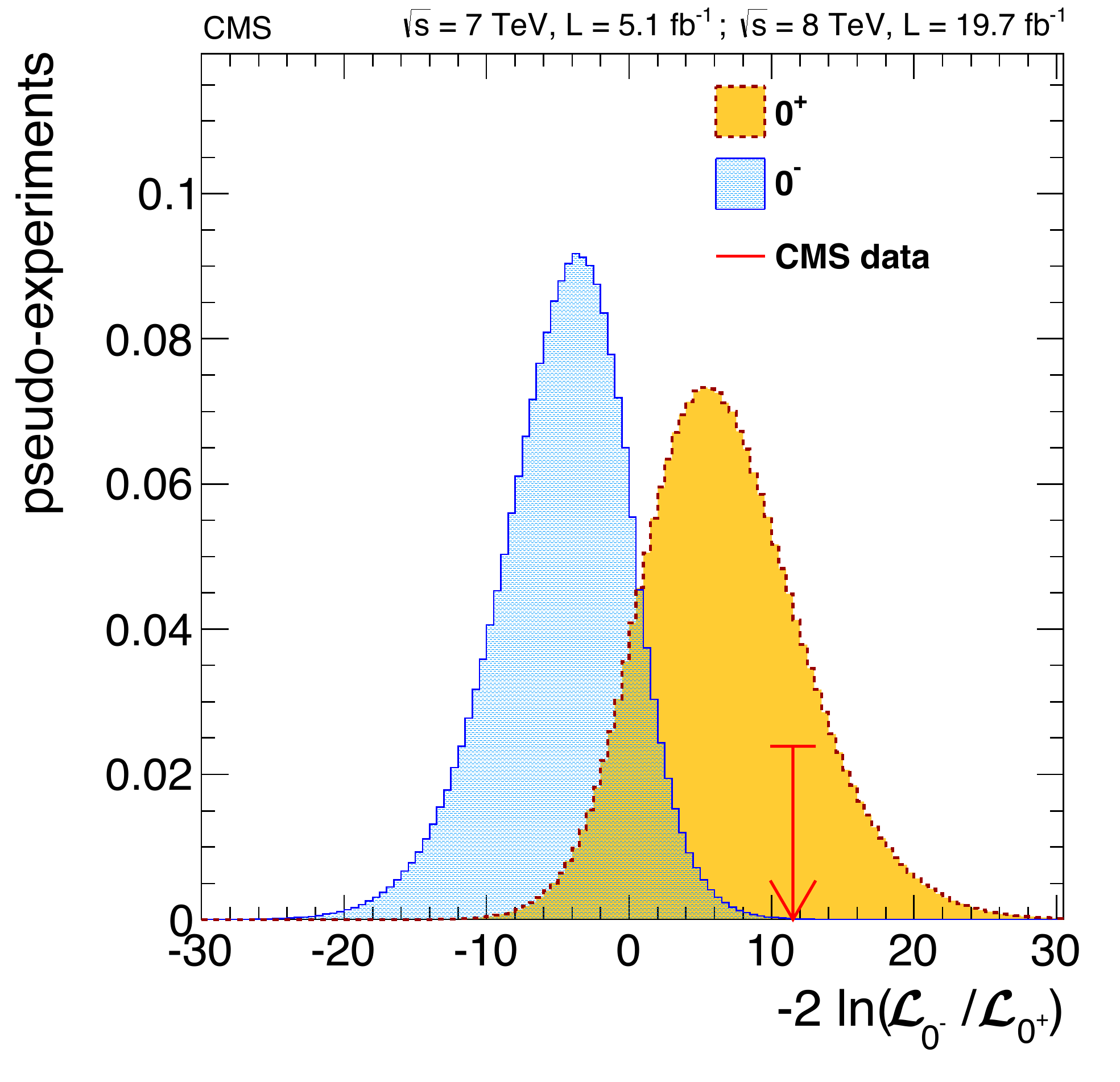}
\includegraphics[height=7cm]{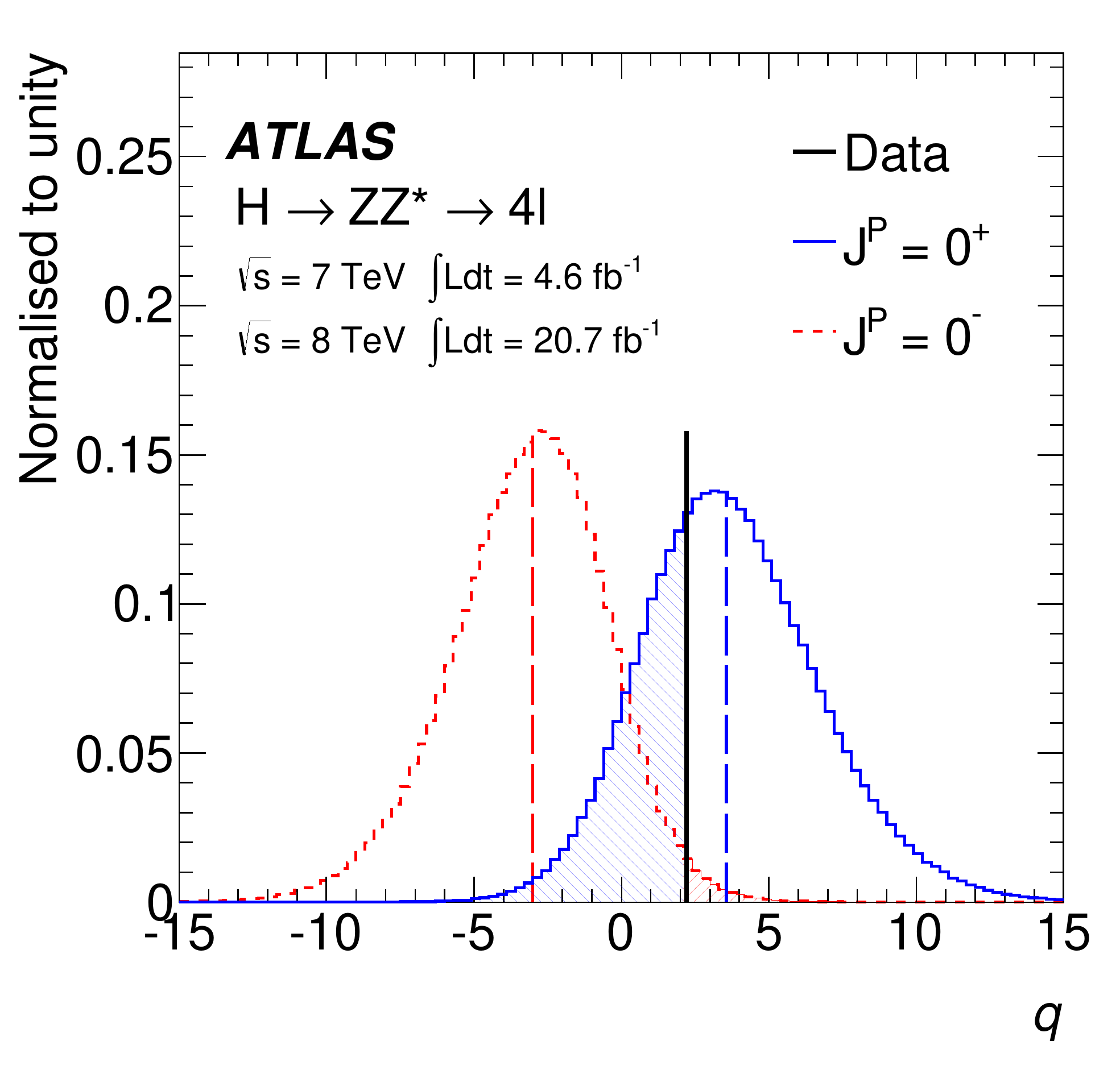}
\caption{\label{fig:SpinMeasurement}Test-statistic distributions for spin parity ($J^P$) hypotheses $0^+$ (SM) and $0^-$ in the $H\rightarrow ZZ^* \rightarrow 4\ell$ channel in ATLAS~\cite{CERN-PH-EP-2013-102} and CMS~\cite{CMS-PAS-HIG-13-002}. The CL${_s}$ probability used to reject the $0^-$ hypothesis is measured as the ratio between the upper tail of the $0^-$ distribution divided by the lower tail of the $0^+$ distribution at the point given by the data. In both experiments the $0^-$ hypothesis is discarded above 99\% probability.}
\end{figure}

In the SM the Higgs boson has a spin parity value of $J^P=0^+$ 
and different alternate hypotheses, namely ($0^-,1^+,1^-,2^+$), are tested against it. 
In the SM the spin 1 is largely suppressed because of the Landau-Yang theorem~\cite{LandauYangTheorem} due to the observation of $H\rightarrow\gamma\gamma$ events, but it could still have other spin values. 
For this, a CL$_s$ probability, as shown in equation~\ref{eq:Cls}, measures the odds of the alternate spin parity hypothesis to the SM hypothesis. 
All hypotheses are rejected at more than 97\% in ATLAS and CMS, which indicates that if the observed boson is not the SM Higgs boson it must deviate very little from it.
Figure~\ref{fig:SpinMeasurement} shows the distributions of the test-statistic for the $0^+$ and $0^-$ hypothesis. 
In this case, the alternate hypothesis $0^-$ is rejected with a confidence level above 99\%.

\begin{equation}
CL_{s}(J^P_\text{alt}) = \frac{p_0(J^P_\text{alt})}{1-p_0(0^+)}
\label{eq:Cls}
\end{equation}

\begin{figure}[ht!]
\centering
\includegraphics[height=6.5cm, trim=0 -30 15 0, clip]{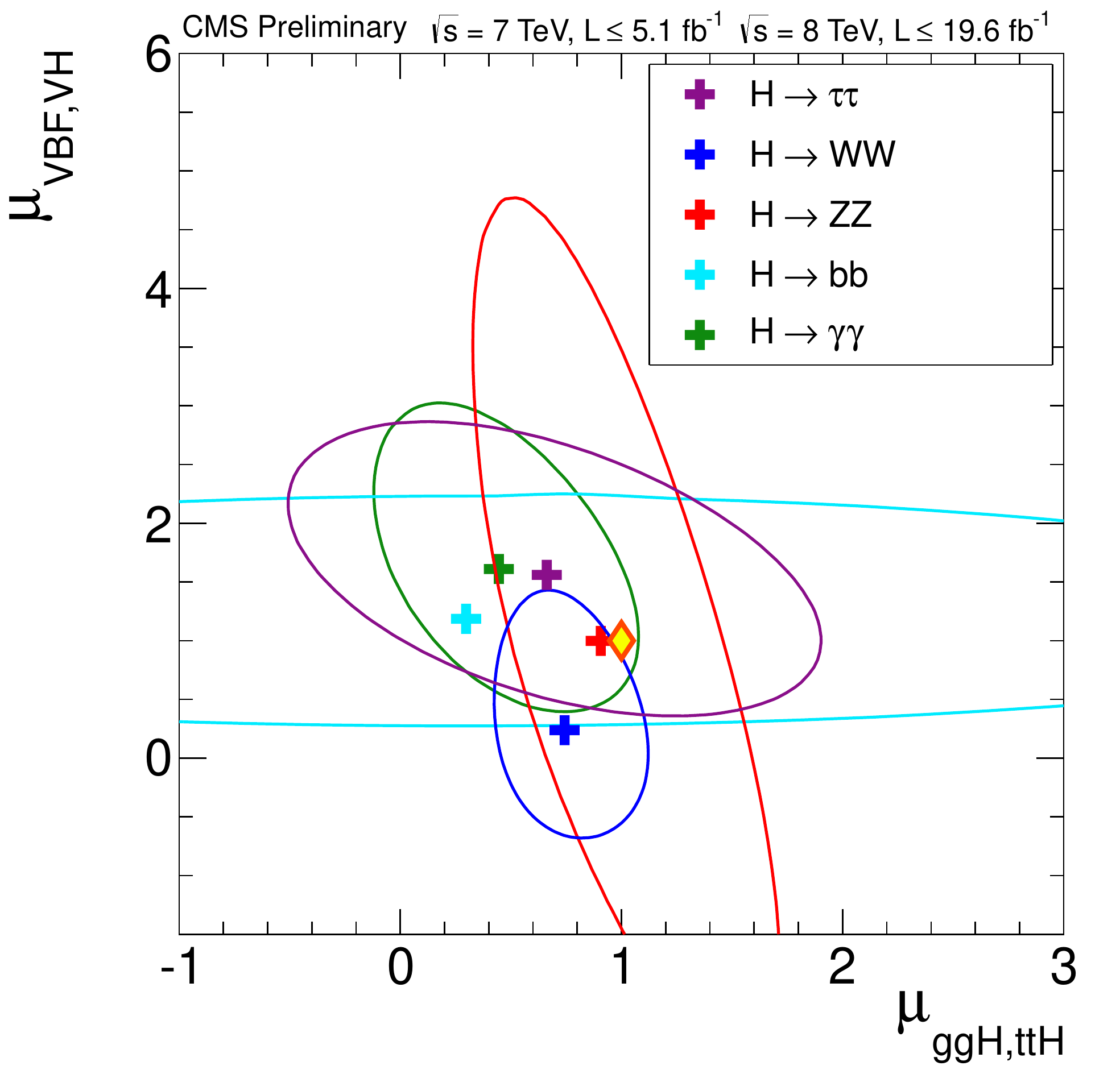}
\includegraphics[height=6.7cm, trim=20 0 25 -10, clip]{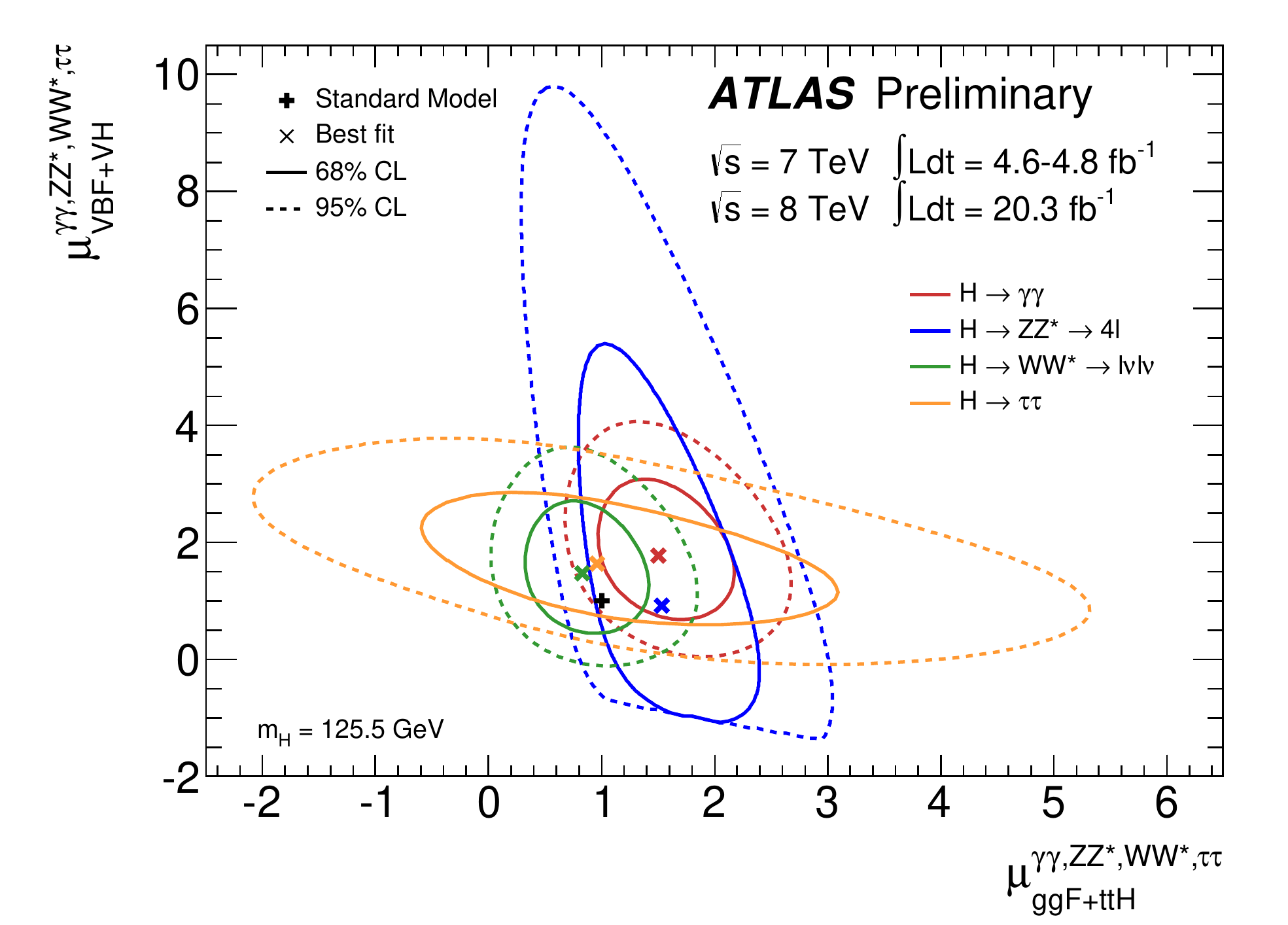}
\caption{\label{fig:VbfProduction}Signal strength for VBF and VH production versus ggH and ttH production contour plots for ATLAS~\cite{ATLAS-CONF-2014-009} and CMS~\cite{CMS-PAS-HIG-13-005}.}
\end{figure}

It is also possible to probe the existence of Higgs production via electroweak processes in a model independent way by measuring the ratio of the gluon fusion ($ggH$) and top fusion ($ttH$) production signal strength to the vector boson fusion (VBF) and associated production (VH) signal strength. This ratio is $\mu_{VBF+VH}/\mu_{ggF+ttH} = 1.4^{+0.5}_{-0.4}(\text{stat})^{+0.4}_{-0.2}(\text{sys})$ for ATLAS and $\mu_{VBF+VH}/μ_{ggF+ttH} = 1.538^{+1.611}_{-0.743}$ for CMS, which corresponds to an evidence of VBF production at the level of 4.1$\sigma$ in ATLAS and 3.21$\sigma$ in CMS.
Figure~\ref{fig:VbfProduction} shows the 68\% probability contour plots of $\mu_{VBF+VH}$ versus $\mu_{ggF+ttH}$, where the SM Higgs boson point $\mu_{VBF+VH}$ = $\mu_{ggF+ttH}$ = 1 is compatible with all measurements.

\section{Higgs couplings}

\begin{figure}[ht!]
\centering
\includegraphics[height=8.7cm,trim=60 0 10 -20,clip]{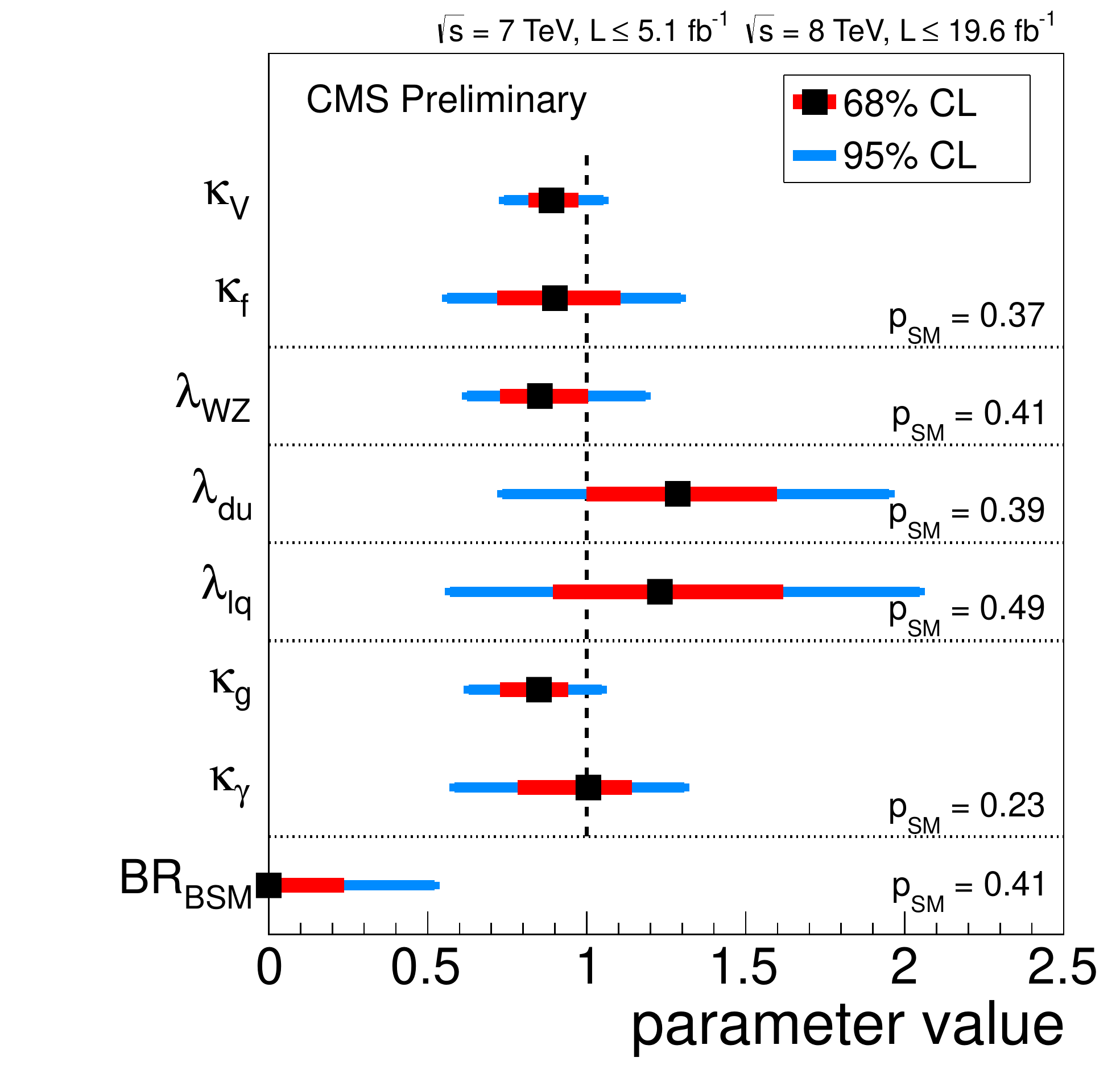}
\includegraphics[height=9.6cm,trim=0 0 0 0,clip]{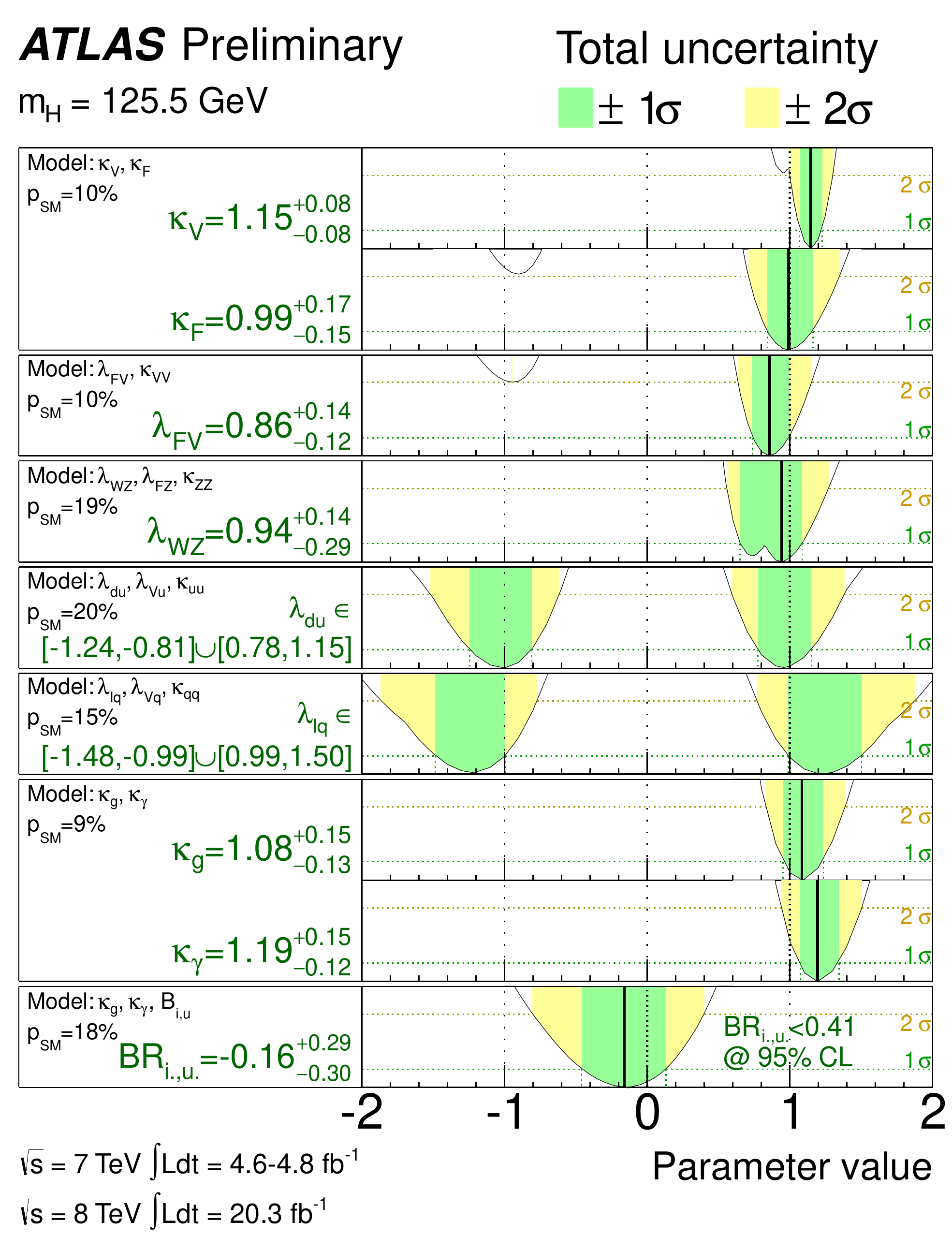}
\caption{\label{fig:Couplings}Higgs boson couplings scale factors in different couplings parametrizations for ATLAS~\cite{ATLAS-CONF-2014-009} and CMS~\cite{CMS-PAS-HIG-13-005} experiments.}
\end{figure}

Following the leading order tree level motivated framework described by the LHC Higgs Cross Section Working Group~\cite{YellowReport2013}, 
measurements of coupling scale factors are implemented for a set of benchmark scenarios.
In these measurements it is assumed that the signal observed originates from a single narrow resonance with the mass measured by the corresponding experiment, a negligible width, and a CP-even state. 
Hence the product of the cross section times the branching ratio for an initial state $i$, a partial decay width into final state $f$ of $\Gamma_f$, and a total decay width of the Higgs boson $\Gamma_H$ as the sum of all visible and invisible final states, is given by

\begin{equation}
\sigma \times \text{BR}(i\rightarrow H \rightarrow f) = 
\frac{\sigma_i \cdot \Gamma_f}{\Gamma_{\text{H}}}\text{.}\nonumber
\end{equation}
Scale factors $\kappa_i$ are added to each coupling and are fitted to the data to test for the modification of the magnitude of the coupling but not its tensor structure. 
In this framework, each final state can involve more than one coupling.
The results for individual channels and their combination are shown in Figure~\ref{fig:Couplings}. 

\begin{figure}[ht!]
\centering
\includegraphics[height=6.5cm,trim=0 -30 0 0,clip]{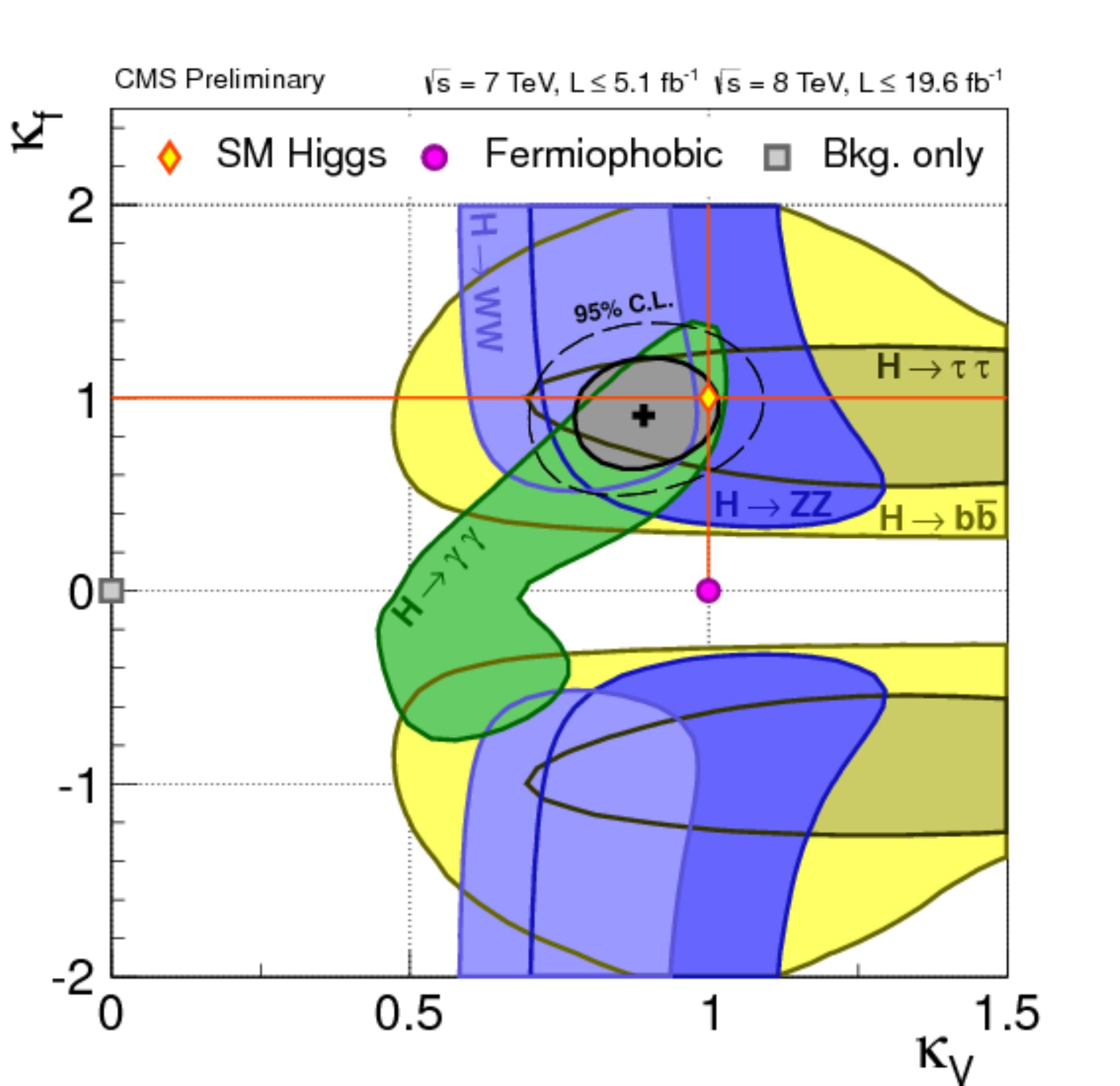}
\includegraphics[height=6.5cm,trim=0 0 0 -10,clip]{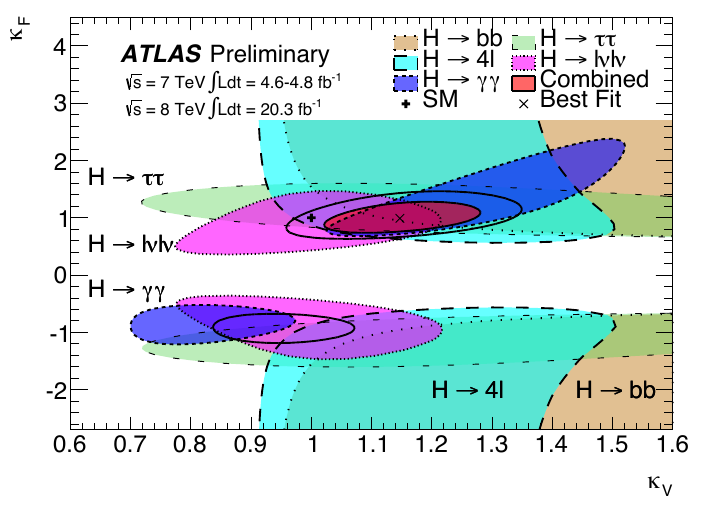}
\caption{\label{fig:CouplingsFV}Higgs boson couplings scale factors to fermions ($\kappa_F$) versus vector bosons ($\kappa_V$) and 68\% CL contours assuming only SM contributions to the total width derived from the individual channels and their combination for ATLAS~\cite{ATLAS-CONF-2014-009} and CMS~\cite{CMS-Higgs-Physics-Results}.}
\end{figure}

Fermion versus vector boson scale factor couplings are tested assuming that only SM particles contribute to the total width. 
Figure~\ref{fig:CouplingsFV} shows the preferred values of coupling scaling factors to fermions ($\kappa_F$) and vector bosons ($\kappa_V$) in a two dimensional fit. The best fit values are are found to be compatible with the SM expectation at the level of 10\%, with a largely constrained $\kappa_F$ due to the observation of the coupling to fermions in the $H\rightarrow\tau\tau$ channel at the level of 4$\sigma$.
When no assumption is made to the total width the measurements in ATLAS are consistent with large signal strength in bosonic decays.

The custodial symmetry in SU(2) that keeps $\rho = m^2_W/m^2_Z \cdot cos^2 \theta_W \approx 1$ is tested by measuring the ratio of the coupling scale factors $\lambda_{WZ} = \kappa_W/\kappa_Z$. The measurement is compatible with the SM prediction and even when an effective scale factor ratio is added to $H\rightarrow\gamma\gamma$ to account for possible contributions beyond the SM.

Up and down type fermion symmetry which is of interest for two Higgs doublet models is probed by the ratio of up to down coupling scale factors $\lambda_{ud} = \kappa_u / \kappa_d$. This measurement provides a 3.6~$\sigma$ evidence of the coupling of the Higgs boson to down type fermions mostly coming from the $H\rightarrow\tau\tau$ measurement. Similarly, quark and lepton symmetry is probed with the ratio $\lambda_{l q} = \kappa_l / \kappa_q$, and a vanishing coupling of the Higgs boson to leptons is excluded at the 4~$\sigma$ level due to the $H\rightarrow\tau\tau$ measurement as in the previous case.

In addition, both experiments probe the contribution of beyond SM particles either in loops or in new final states by introducing effective coupling scale factors for the $H\rightarrow gg$ and $H\rightarrow\gamma\gamma$ vertices, $\kappa_g$ and $\kappa_\gamma$, and assuming the rest of the coupling scale factors to be as predicted by the SM. These vertices are very sensitive to unknown heavy particles beyond the SM. 
In the first benchmark model it is assumed that there are no sizeable contributions to the total width by non-SM particles and the free parameters are $\kappa_g$ and $\kappa_\gamma$. 
These measurements show the lowest compatibility with the SM in ATLAS due to the high signal strength value observed.

Finally, upper limits to the branching ratio to invisible or undetected final states can be set by considering $BR_{BSM} = \Gamma_{BSM}/\Gamma_{H} = 1 - \kappa^2_H \cdot \Gamma^{SM}_H / \Gamma_H$. CMS quotes $BR_{BSM} < $~0.52 at 95\% CL and ATLAS quotes $BR_{BSM} <$~0.41 for the same CL with a noticeable difference over the expected value ($BR_{BSM} < $~0.55) mostly constrained from channels sensitive to VBF and VH production like $H\rightarrow b\bar{b}$ and $H\rightarrow \tau\tau$.

\section{Limits on new physics}

\begin{figure}[ht!]
\centering
\begin{subfigure}{0.49\textwidth}
	\includegraphics[height=5.5cm,trim=30 10 0 0,clip]{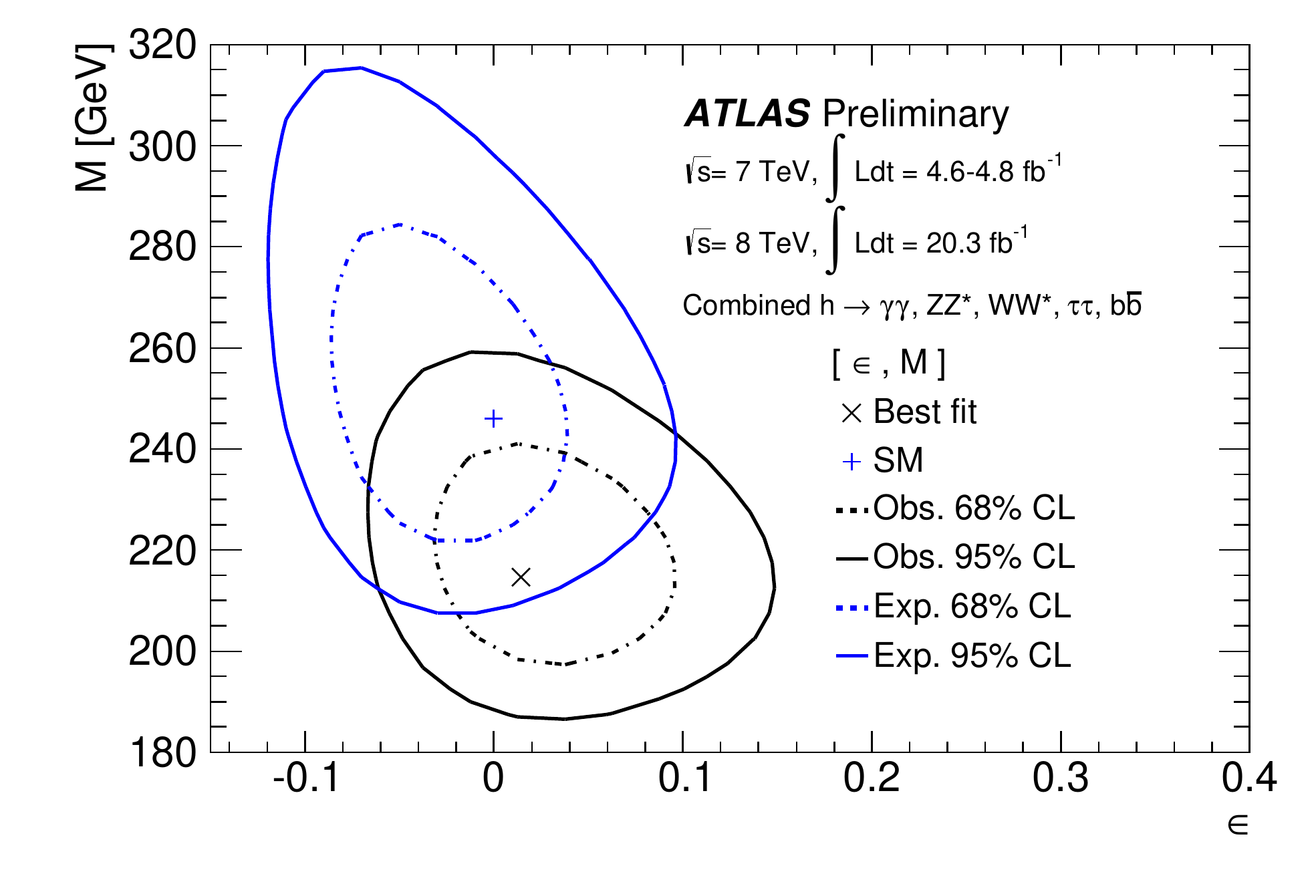}
	\caption{\label{fig:MassScaling}}
\end{subfigure}
\hfill
\begin{subfigure}{0.49\textwidth}
	\includegraphics[height=5.5cm,trim=0 10 0 0,clip]{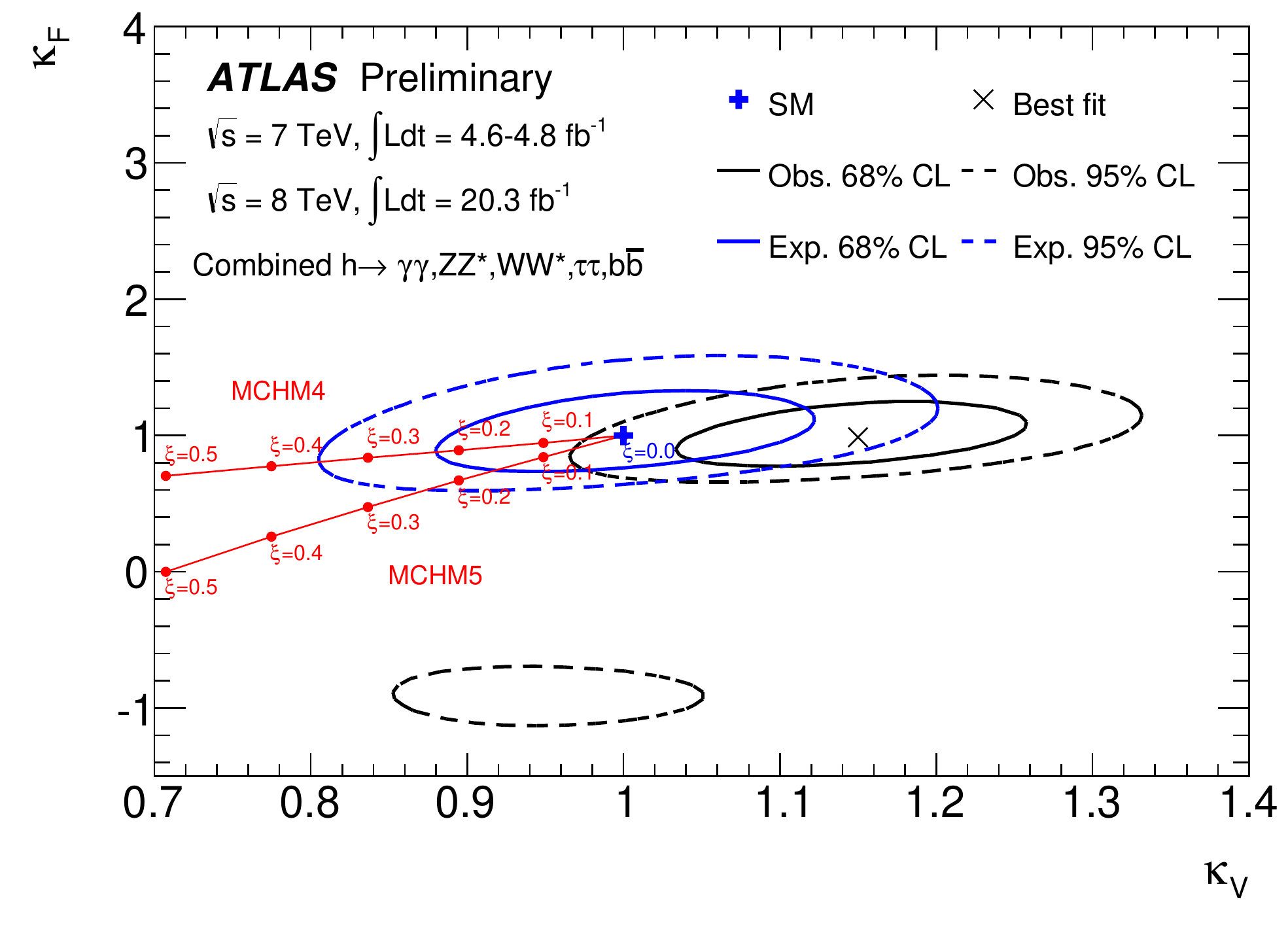}
	\caption{\label{fig:Mchm}}
\end{subfigure}
\caption{(a) Contour plot of the vacuum expectation value ($M$) versus the mass scaling paramter ($\varepsilon$) in ATLAS~\cite{ATLAS-CONF-2014-010}. (b) Contour plot of the coupling scale factors $\kappa_F$ versus $\kappa_V$ in ATLAS~\cite{ATLAS-CONF-2014-010}.}
\end{figure}

The mass dependence of the couplings is probed by expressing the coupling scale factors to different species of fermions and vector bosons in terms of a mass scaling parameter ($\varepsilon$) and a vacuum expectation value ($M$).
%
Combined fits to measured rates are performed as a function of $\varepsilon$ and $M$, and the best fit point is compatible with the SM Higgs boson at the level of 1.5~$\sigma$, as it is shown in Figure~\ref{fig:MassScaling}.
The best fit for the vacuum expectation value is below the one for the SM because the overall signal strength is higher than 1.

If the Higgs were a composite particle following Mininal Composite Higgs Models (MCHM), the couplings to fermions and vector bosons could be modified by its compositeness scale parameter ($f$) such as $\xi = v^2/f^2$, and the SM Higgs boson is recovered in the limit $\xi\rightarrow 0$ and $f\rightarrow \infty$.
In particular, in the model MCHM4, the couplings scale factors are $\kappa = \kappa_F = \kappa_V = \sqrt{1-\xi}$ and in the MCHM5 $\kappa_V = \sqrt{1-\xi}$ and $\kappa_F = \frac{1-2\xi}{\sqrt{1-\xi}}$.
Figure~\ref{fig:Mchm} shows the two dimensional likelihood contours in the ($\kappa_V$ , $\kappa_F$ ) coupling plane, and the coupling predictions in the MCHM4 and MCHM5 models as parametric functions of the Higgs boson compositeness parameter $\varepsilon$. 
The upper limit of the composinteness scale is $f$ $>$ 710 GeV for MHCM4 and $f$ $>$ 640 GeV for MHCM5. 

\begin{figure}[ht!]
\centering
\includegraphics[height=6cm,trim=0 10 0 10,clip]{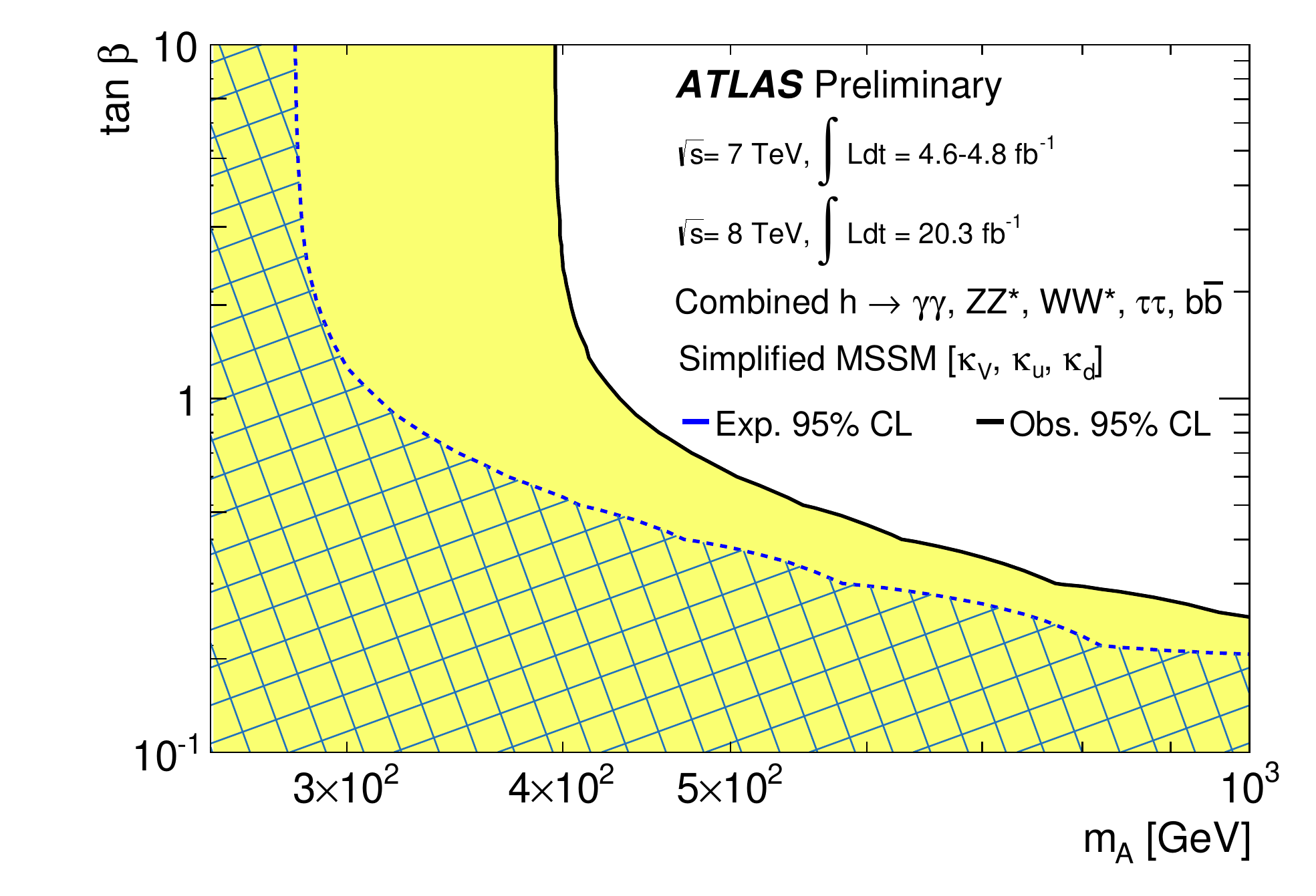}
\caption{\label{fig:LimitsMssm}Exclusion region at 95\% CL in the ($m_A$,$\tan\beta$) plane in a simplified MSSM model via a fit to the measured rates of the Higgs boson production and decays for ATLAS~\cite{ATLAS-CONF-2014-010}.}
\end{figure}

In the two Higgs Doublet Model (2HDM) where the SM Higgs sector is extended by an additional doublet predicting the existence of five Higgs bosons, two neutral CP-even, one neutral CP-odd and tho charged bosons, with a vacuum expectation value that follows the relation $v^2_1 + v^2_2 = v^2 \approx (246 \text{ GeV})^2$, the type II category is the most interesting among the four that exist depending on their coupling constants. 
In the type II 2HDM where couplings are different for up-type quarks and for down-type quarks and leptons, and are completely determined by the mass of the CP-odd scalar ($m_A$) and the ratio between the up and down type fermions ($\tan\beta=v_u/v_d$) results can be interpreted in a simplified MSSM model. In this model the loop corrections from stops in $ggF$ production and $\gamma\gamma$ decays are ignored and the Higgs boson decays to supersymmetric particles are neglected.
The measured production and decay rates are expressed in terms of the corresponding couplings, assuming identical production to the SM but without any assumption on the total Higgs width. The measurements are shown in Figure~\ref{fig:LimitsMssm}.
The observed exclusion limit at 95\% CL is stronger than expected since measured rates in $H\rightarrow \gamma\gamma$ and $H\rightarrow ZZ^*\rightarrow 4\ell$ are higher than predicted. 

\begin{figure}[ht!]
\centering
\includegraphics[height=6.4cm,trim=8 0 5 0,clip]{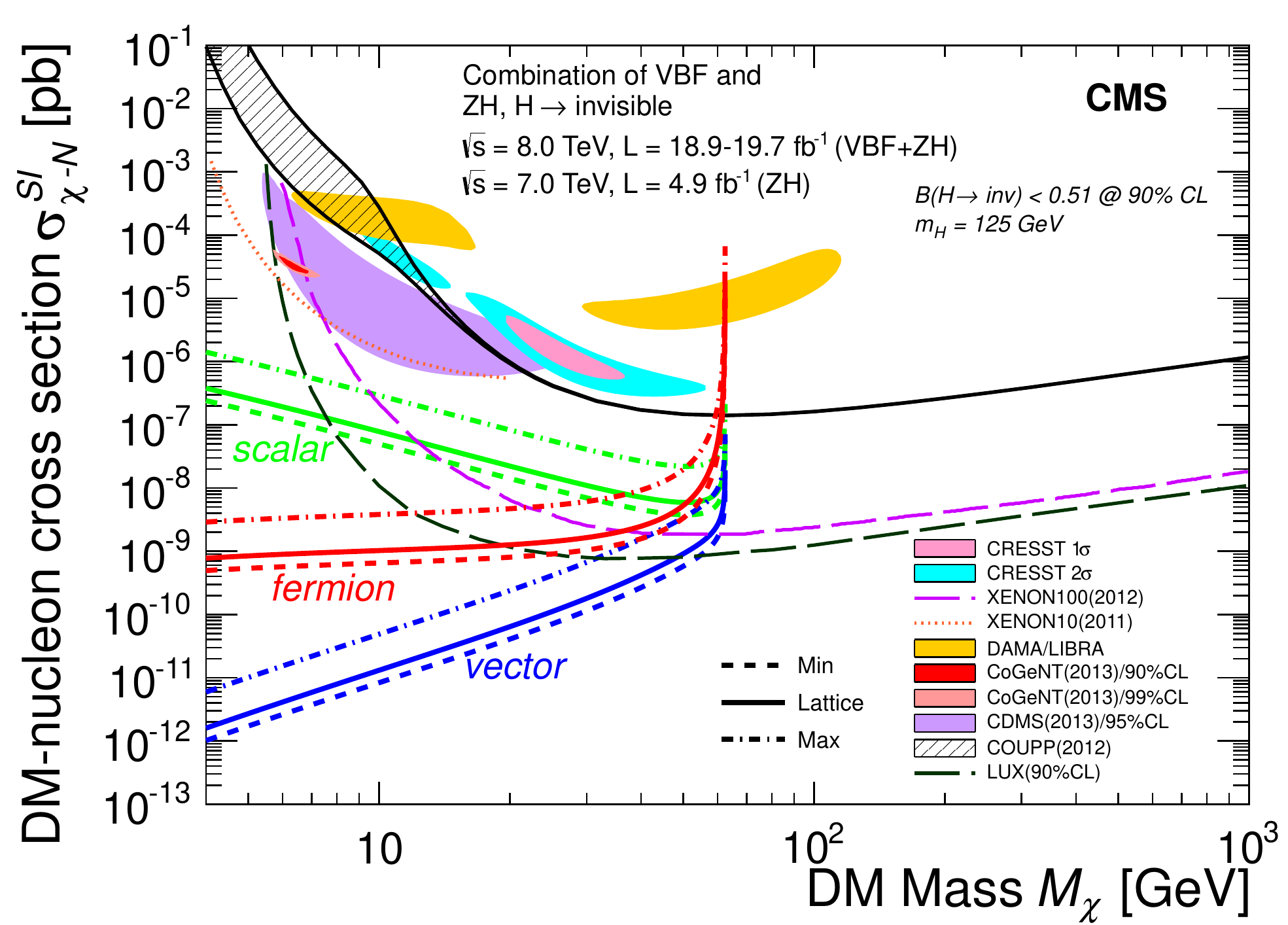}
\includegraphics[height=6.4cm,trim=10 0 15 0,clip]{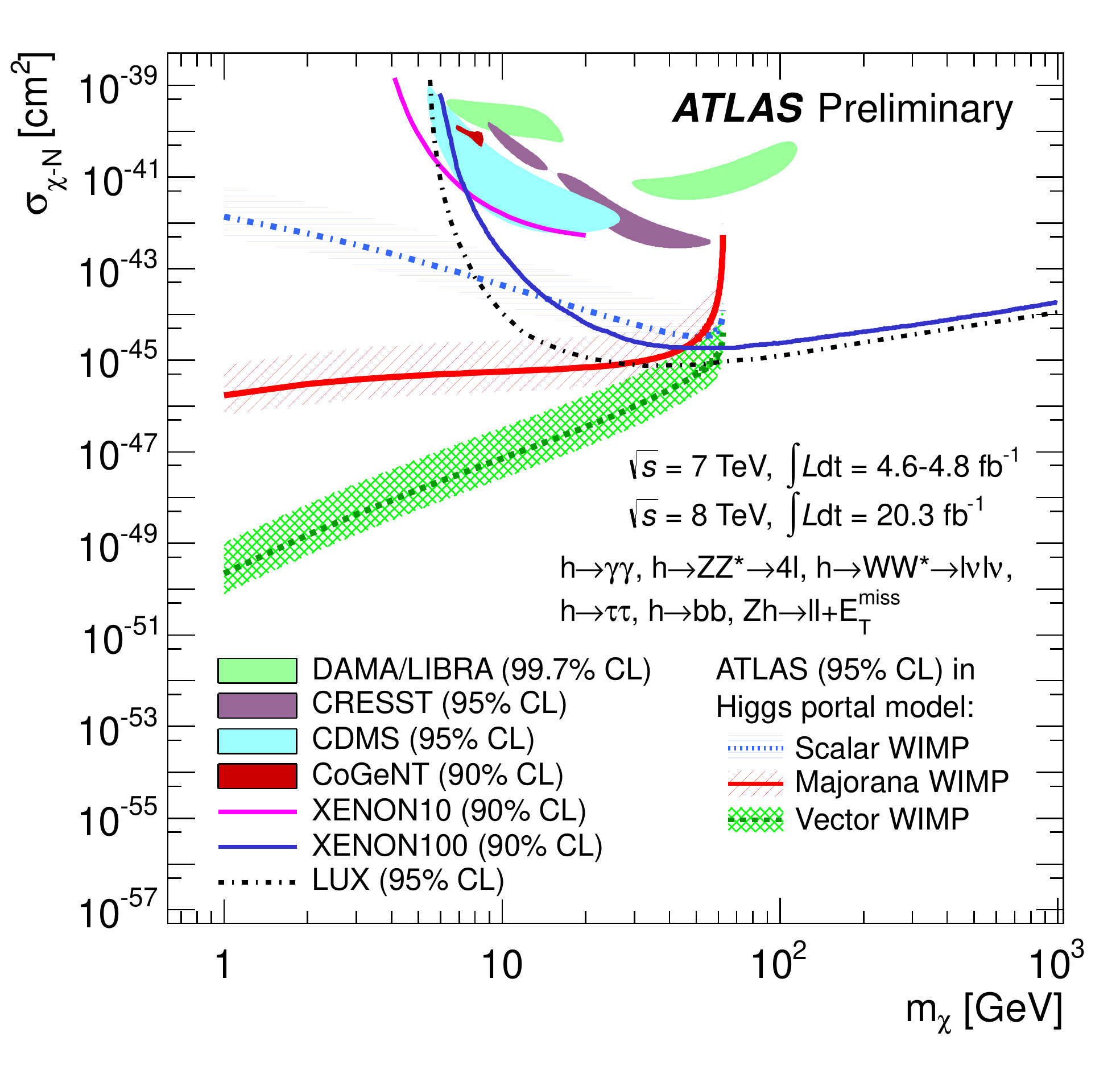}
\caption{\label{fig:LimitsDarkMatter}Upper limits at 95\% CL on the WIMP-nucleon scattering cross section as a function of the WIMP mass $m_\xi$ (scalar, vector or fermion) compared with the limits from direct searches for dark matter at the confidence levels indicated for ATLAS~\cite{ATLAS-CONF-2014-010} and CMS~\cite{CMS-PAS-HIG-13-030}.}
\end{figure}

It is conceivable that the Higgs particle may have other decay channels that are not predicted by the SM.
In a wide context, the Higgs boson could be coupled to the particle that constitutes all or part of the dark matter in the universe. These are so called Higgs portal models.
The upper limit on the branching ratio of the Higgs boson to invisible final states is derived using the combination of rate measurements from the individual channels and the measured upper limit on the rate of the $Zh\rightarrow\ell\ell + E_T^{miss}$ process, proportional to $BR_{BSM}$. The couplings of the Higgs boson to massive particles other than Weak Interacting Massive Particles (WIMPs) are assumed to be equal to the SM predictions, allowing the corresponding partial decay widths and invisible decay width to be inferred.
Limits are considerably more stringent at low mass and degrade as $m_X$ approaches $m_H/2$ as shown on Figure~\ref{fig:LimitsDarkMatter}.

\section{Conclusions and outlook}
ATLAS and CMS discovered a Higgs like particle with a mass close to 125.5~GeV, and measured the spin, parity and signal strength to be compatible with the one from the SM Higgs boson ($J^P=0^+$ , $\mu_{VBF+VH}$ = $\mu_{ggH+ttH}$ = 1).
In a coupling scale factors analysis, compatibility with the SM is found in all the tests performed, with probabilities ranging from 7\% to 21\%. 
Consequently, the Higgs physics potential of the LHC Run-I is almost exploited.
Run-II and beyond will offer the possibility to measure the couplings more precisely, further constrain rare decays, and determine a possible CP admixture of the Higgs boson.


\end{document}